\documentclass[5p]{elsarticle}

\include{graphicx.sty}
\usepackage{amsmath}
\journal{Physics Letters B}

\begin{document}

\begin{frontmatter}

\title{Neutrino decoherence in presence of strong gravitational fields}

\author{Am\'elie Chatelain and Maria Cristina Volpe}
\ead{chatelai@apc.in2p3.fr,volpe@apc.univ-paris7.fr}
\address{Astro-Particule et Cosmologie (APC), CNRS UMR 7164, Universit\'e Denis Diderot,\\ 10, rue Alice Domon et L\'eonie Duquet, 75205 Paris Cedex 13, France}

\begin{abstract}
We explore the impact of strong gravitational fields on neutrino decoherence. To this aim,
we employ the density matrix formalism 
to describe the propagation of neutrino wave packets in curved spacetime. 
By considering Gaussian wave packets, we determine the coherence proper time, neglecting the effect of matter outside the compact object. We show that strong gravitational fields nearby compact objects  significantly influence neutrino coherence.

\end{abstract}


\end{frontmatter}

\section{Introduction}
\label{intro}
\noindent
Neutrinos are elementary massive particles with non-zero mixings producing neutrino oscillations, a quantum mechanical phenomenon
analogous to Rabi oscillations in ato\-mic physics \cite{Fukuda:1998mi,Ahmad:2002jz}.
This phenomenon depends on the fact that the flavor and the mass basis are
related by the Pon\-te\-cor\-vo-Maki-Nakagawa-Sakata unitary matrix (PMNS) who\-se 
mixing angles are known. As for the Dirac CP violating phase, there are indications for it to be large \cite{Capozzi:2018ubv}, while Majorana phases remain unknown.
Neutrino masses require extensions 
of the Glashow-Weinberg-Salam Model as e.g. including three right handed singlets and Yukawa couplings. 

Neutrino oscillation studies typically employ the plane wave approximation  to describe flavor conversion.  
Wave packets (WPs) account for neutrinos being localized particles. 
Their use introduces
decoherence among the mass eigenstates 
due to the finite extension of corresponding WPs \cite{Nussinov:1976uw}.
In laboratory experiments, the WP widths include the finite size both at neutrino production and at detection. In the WP treatment an exponential factor suppresses coherence in the interference term of the oscillation probabilities.  
However, at typical distances of oscillation
experiments, this correction is negligeable \cite{Giunti:2003ax,Giunti:2007ry,An:2016pvi}. 
Besides WP separation, other mechanisms produce
neutrino decoherence, such as the propagation in a quantum gravity foam
\cite{Barenboim:2006xt}, for which experimental constraints exist (see e.g. \cite{Fogli:2007tx}). 

In matter the WP widths associated
with neutrino production processes (such as inverse beta-decay) are small, around $10^{-11}-10^{-12}~$cm 
\cite{Kersten:2015kio,Akhmedov:2017mcc}. 
In dense environments the WP treatment has implications depending on the adiabaticity.
In case of adiabatic evolution, a WP description in the matter basis introduces
an exponential suppression factor with 
a coherence length that is similar to the vacuum case \cite{Kersten:2015kio,Akhmedov:2017mcc}.
On the other hand, if neutrino evolution is non-adiabatic, mass eigenstate mixing is not suppressed which makes difficult to even define a coherence length. Consequently decoherence by WP separation depends on the model chosen for adiabaticity violation \cite{Akhmedov:2017mcc}. 

So far, investigations of neutrino flavor conversion based on WPs have been performed in flat spacetime.
Nearby a neutron star or a black hole, strong gravitational fields influence 
neutrino flavor evolution. Effects due to trajectory bending on nucleosynthetic outcomes have been studied in a black hole accretion disk \cite{Caballero:2011dw}, while Ref. \cite{Deaton:2018ser} has presented a general relativistic ray tracing for neutrinos.
Neutrino trajectories in the Schwarzschild metric are explored in \cite{Fornengo:1996ef}. Ref.\cite{MosqueraCuesta:2017iln} presents the neutrino dynamics and the influence on the electron fraction of a slowly rotating nonlinear charged black hole.
In dense environments, gravitational fields can also modify the localization and adiabaticity of the Mikheev-Smirnov-Wolfenstein resonance \cite{Cardall:1996cd},
or delay bipolar oscillations from neutrino self-interactions \cite{Yang:2017asl}.
Such studies are based on the plane wave approximation.
 
In this letter we explore the impact of strong gravitational fields 
on neutrino decoherence in a WP treatment.
We first recall the density matrix formalism using WPs in the case of flat spacetime. Then
we extend it to describe neutrino flavor evolution 
in curved spacetime, considering a static and spherically symmetric gravitational field
described by the Schwarzschild metric. 
Neutrino decoherence in curved spacetime is quantified by a coherence proper time, instead of a coherence proper length, as in flat spacetime.
We first introduce kinematical arguments and then derive the neutrino coherence proper time 
based on density matrices with Gaussian WPs.
We neglect matter and neutrino self-interactions outside the compact object.
Finally, we present numerical estimates of the gravitational field effects on the coherence
proper time.  

This letter is structured as follows. 
Section II recalls the WP treatment of neutrino oscillations in vacuum and introduces the coherence length in flat spacetime.
Section III presents the extension of the formalism to curved spacetime in the Schwarzschild metric. 
Then kinematical arguments are presented and the derivation of the coherence proper time based on the density matrix formalism for WPs. Numerical results for the coherence proper time are shown.  Section IV gives our conclusions.

\section{Neutrino WP decoherence in flat spacetime}
\label{flat spacetime}
\subsection{Neutrino states}
\noindent
A neutrino state in coordinate space can be Fourier expanded as (we take $\hbar = c= G = 1$)\footnote{For brevity we have introduced the shortened notation 
\begin{equation}\label{e:sn}
\int_{\vec{p}} \equiv \int {{d^3 p}\over{(2 \pi)^3}}. \nonumber
\end{equation}}   \cite{Akhmedov:2017mcc} 
\begin{equation}\label{e:sv}
 | \nu (t, \vec{x}) \rangle = \int_{\vec{p}}  e^{i\vec{p}\cdot\vec{x}}  |  \nu (t, \vec{p})  \rangle,
\end{equation}
with $ | \nu (t, \vec{p}) \rangle$ the time-dependent state  with momentum 
$\vec{p}$. 
A vector of N such states, $N$ being the number of neutrino families, is solution of the Schr\"odinger-like equation 
\begin{equation}\label{e:ev}
i {{\rm d}\over{\rm dt}} | \nu (t, \vec{p}) \rangle = h(t,\vec{p})  ~ | \nu (t, \vec{p}) \rangle,
\end{equation} 
where $h(t,\vec{p})$ is the Hamiltonian governing neutrino evolution.
In astrophysical environments, it includes different contributions
\begin{equation}\label{e:h}
h(t, \vec{p}) = h_{\text{vac}}(\vec{p}) + h_{mat}(t) + h_{\nu\nu}(t),
\end{equation}
where the first is the vacuum term, the second and the third are the mean-field contributions from neutrino interactions with matter and with (anti)neutrinos respectively. In fact, 
neutrino self-interactions give sizeable effects in dense media 
such as core-collapse supernovae or binary neutron star merger remnants \cite{Duan:2010bg}. 
The vacuum term is $h_{\text{vac}}(\vec{p}) = U h_{0}(\vec{p}) U^{\dagger}$, with  $h_{0} = \text{diag}(E_j(\vec{p}))$. The quantity $E_j(\vec{p}) = \sqrt{\vec{p}^2 + m_j^2}$ is
the energy eigenvalue of the $j$th mass eigenstate, with $j \in [1,{\rm N}]$.

At each time,
a neutrino flavor state is a superposition of the mass eigenstates  
\begin{equation}\label{e:sup}
 | \nu_{\alpha} (t, \vec{p}) \rangle = U^*_{\alpha j}  | \nu_j(t, \vec{p})\rangle,
\end{equation}
where $\alpha$ stands for flavor.
The quantity $U$ is the Pontecorvo-Maki-Nakagawa-Sakata  unitary matrix relating the flavor to the mass basis \cite{Tanabashi:2018oca}. In three flavors, it depends on three mixing angles and three CP violating phases (one Dirac and two Majorana).

Usually, treatments of flavor conversion consider the neutrino mass eigenstates as plane waves and that the neutrino flavor state satisfies the light-ray approximation, i.e. $L=t$, with $L$ the travelled distance. In a WP description, a neutrino flavor state Eq.\eqref{e:sup} becomes a superposition of the mass eigenstates WPs. Each momentum component satisfies Eq.\eqref{e:ev} as far as the size of the momentum distribution is large, compared to the inverse length beyond which the interaction potentials vary. In the present work, we neglect the presence of matter and neutrino self-interactions outside the compact object. Therefore, from now on, we only keep the vacuum term in Eq.\eqref{e:h}.

At initial time each WP component satisfies
\begin{equation}\label{e:wp}
 |  \nu_j(t_0, \vec{p})\rangle = f_{\vec{p}_j}(\vec{p})   | \nu_j^{(0)}(t_0,\vec{p})\rangle,
\end{equation}
where $ | \nu_j^{(0)}(t_0,\vec{p})\rangle$ are propagation eigenstates satisfying
\begin{equation}
\langle  \nu_k^{(0)}(t_0,\vec{p}~') | \nu_j^{(0)}(t_0,\vec{p})\rangle = (2 \pi)^3 \delta(\vec{p}~' - \vec{p}) \delta_{jk}.
\end{equation}
The quantities $f_{\vec{p}_j}(\vec{p})$ are the momentum distribution amplitudes centered 
at momentum $\vec{p}_j$ which describe the WP associated to the $j$th eigenstate of mass $m_j$. They are normalised according to
\begin{equation}\label{eq:Gnor}
\int_{\vec{p}}  \left| f_{\vec{p}_j} (\vec{p}) \right|^2 = 1.
\end{equation}
 
The neutrino flavor state in coordinate space can be written as 
\begin{equation}
 | \nu (t, \vec{x}) \rangle = U^*_{\alpha j}  \psi_j(t, \vec{x})  |  \nu_j \rangle,
\label{eq:nus}
\end{equation}
where the coordinate-space wave function of the $j$th mass eigenstate is related to the momentum dependent wave function according to
\begin{equation}
\psi_j (t, \vec{x})= \int_{\vec{p}}  e^{i\vec{p}\cdot\vec{x}} \psi_j (t, \vec{p}).
\label{eq:psi}
\end{equation}
From Eqs.(\ref{e:ev}-\ref{e:h}) 
(where, in Eq.\eqref{e:h}, $h_{mat}$ and $h_{\nu\nu}$ are discarded) and \eqref{e:wp} the time evolution of its Fourier components follows
\begin{equation}
\psi_j (t, \vec{p}) = f_{\vec{p}_j}(\vec{p})  e^{-i E_j  (\vec{p})) t}.
\label{eq:psit}
\end{equation}

In our investigation we employ the density matrix formalism to describe
neutrino WPs decoherence. For flat spacetime we follow the derivation performed in Ref.\cite{Akhmedov:2017mcc}\footnote{Refs.\cite{Giunti:2003ax,Giunti:2007ry} give an alternative approach using neutrino amplitudes.}. The
one-body density matrix is given by\footnote{The mean-field approximation for one-body density matrices corresponds to the first truncation of the
Born-Bogoliubov-Green-Kirkwoord-Yvon hierarchy (BBGKY) which is a hierarchy
of equations of motion for reduced many-body density matrices. Ref.\cite{Volpe:2013jgr} has applied its relativistic generalisation to a system of neutrinos and antineutrinos, as plane waves, propagating in an astrophysical environment.} 
\begin{equation}
\rho (t, \vec{x}) =   | \nu (t, \vec{x}) \rangle \langle \nu (t, \vec{x})  |,
\label{eq:oneb}
\end{equation}
with a similar expression for antineutrinos\footnote{Note that the same evolution equations hold for neutrinos  and antineutrinos, by taking $\rho_{ij} = \langle a^{\dagger}_j a_i \rangle$ and $\bar{\rho}_{ij} = \langle b^{\dagger}_i b_j \rangle$ respectively. The creation $a^{\dagger}$ ($b^{\dagger}$) and annihilation $a$ ($b$) operators for neutrinos (antineutrinos) satisfy the equal time canonical commutation rules \cite{Volpe:2013jgr}.}.
Their evolution is governed by the Liouville Von Neumann equation 
\begin{equation}
i D {\rho} = [h, \rho],
\end{equation}
where the $D = {{\partial} \over {\partial t}} + \vec{v}_g {\mathrm d \over{\mathrm d\vec{x}}}$ is Liouville operator, with $\vec{v}_g$ the group velocity of the neutrino WP.
By using Eqs.(\ref{eq:psi}-\ref{eq:oneb}), the $jk$-matrix elements of the one-body density matrix can be written as
\begin{equation}
\rho_{jk} (t, \vec{x}) = U^*_{\alpha j} U_{\alpha k} \psi_j (t, \vec{x}) \psi_k^* (t, \vec{x}),
\label{eq:onebelem}
\end{equation}
where $k$ denotes the $k$th mass eigenstate.
\subsection{Coherence length in flat spacetime}
In a WP treatment the condition for vacuum oscillations to take place is that 
the WPs overlap sufficiently
to produce interference among the mass eigenstates. One defines the coherence length $L_{coh}$ as the distance $L$ at which the separation $\Delta x$ between the mass eigenstates WPs centroids is at least $\sigma_x$ (the WP width), i.e.
\begin{equation}\label{e:gwp}
 L = L_{coh} ~~~~\mathrm{if}~~~~\Delta x = \sigma_x. 
\end{equation}

Heuristically, one can estimate the coherence length as
$L_{coh} \simeq \sigma_x v_g(\Delta v_g)^{-1}$, with $v_{g}$ the average group velocity of the WPs, while $\Delta v_g$ is the difference between the group velocities of the mass eigenstates WP.  
The group velocity for the $j$th mass eigenstate WP is (assuming $m_j/E_j \ll 1$)
\begin{equation}\label{eq:gvel}
v_{j} = \frac{\partial E_j}{\partial  p}\vert_{\vec{p} = \vec{p}_j} \simeq 1 - \frac{m_j^2}{2E^2},
\end{equation}
in vacuum, where $E \simeq \vert \vec{\bar{p}} \vert$ the average energy between the $j$th and the $k$th mass eigenstates. 
Therefore, an heuristic estimate of the coherence length in vacuum is
\begin{equation}\label{e:Lcoh}
L_{coh} = \frac{2 E^2}{\left| \Delta m_{ij}^2 \right|} \sigma_x,
\end{equation}
with $\Delta m_{ij}^2 = m_i^2-m_j^2$. Note that this argument can be extended in the case of neutrinos adiabatic evolution in presence of matter and self-interactions \cite{Akhmedov:2017mcc}. 
\subsection{The density matrix approach}
We now determine the coherence length through the density matrix formalism by considering Gaussian WPs of width $\sigma_p$
\begin{equation}
f_{\vec{p}_j} (\vec{p}) =  (\frac{2\pi}{\sigma_p^2})^{\frac{3}{4}}  \exp{\left[ - \frac{(\vec{p}-\vec{p}_j)^2}{4\sigma_p^2} \right]}.
\label{eq:GWP}
\end{equation}
By using Eqs.(\ref{eq:psi}-\ref{eq:psit}),(\ref{eq:onebelem}),(\ref{eq:GWP}), one gets
\begin{multline}
\rho_{jk} (t, \vec{x}) = N^{\alpha}_{jk}
\int_{\vec{p},\vec{q}}
\exp \left[- i \left[E_j(\vec{p})   - E_k(\vec{q})\right] t \right\rbrace  \\ 
\times \exp{\left[i (\vec{p} - \vec{q})\vec{x} - \frac{(\vec{p}-\vec{p}_j)^2}{4\sigma_p^2} - \frac{(\vec{q}-\vec{p}_k)^2}{4\sigma_p^2} \right]},
\label{eq:exprhojk}
\end{multline}
with the factor
\begin{equation}\label{eq:factor}
N^{\alpha}_{jk} =  (\frac{2\pi}{\sigma_p^2})^{\frac{3}{2}} U^*_{\alpha j} U_{\alpha k}.
\end{equation}
\noindent
To calculate the integrals in \eqref{eq:exprhojk}, we expand the neutrino energies around the peak momenta $\vec{p}_j$, and retain only the first two terms of the expansion
\begin{equation}
E_j (\vec{p}) = E_j  + (\vec{p} - \vec{p}_j) \vec{v}_j 
+ \mathcal{O} \left[(\vec{p} - \vec{p}_j )^2 \right],
\label{eq:expansionE}
\end{equation}
where $E_j \equiv E_j(\vec{p}_j)$ and $\vec{v}_j$ the group velocity of the $j$th mass eigenstate WP \eqref{eq:gvel}. Neglecting higher order terms in the expansion of $E_j (\vec{p})$ amounts to disregarding the WP spread during neutrino propagation. Note that such spread should have no effect on the coherence of supernova neutrinos Ref.\cite{Kersten:2015kio}\footnote{Ref.\cite{Naumov:2013uia} argued that WP dispersion from propagation could induce non-trivial effects.}.

By performing the Gaussian integrals in \eqref{eq:exprhojk}, the matrix elements of the one-body density matrix in coordinate space become
\begin{multline}
\rho_{jk} (t, \vec{x}) =  N^{\alpha}_{jk} {\sigma_p^6 \over {\pi}^3}
\exp{\left[- i (E_{jk}t - \vec{p}_{jk} \vec{x})\right]} \\
\times \exp{\left[- \frac{( \vec{x}-\vec{v}_j t)^2}{4\sigma_x^2} - \frac{(\vec{x}-\vec{v}_k t)^2}{4\sigma_x^2} \right]},
\label{eq:exprhojktx}
\end{multline}
where $E_{jk} \equiv E_j-E_k$, $\vec{p}_{jk} = \vec{p}_j - \vec{p}_k $ and $\sigma_x = (2\sigma_p)^{-1}$ is the neutrino WP size in coordinate space.

In oscillation experiments, the quantity of interest is the neutrino decoherence as a function of distance. By integrating over time Eq.(\ref{eq:exprhojktx})\footnote{Since the WP amplitudes decrease quickly as $t$ deviates from the stationary point of the exponent $t_\text{stat} = \frac{\vec{v}_j+\vec{v}_k}{\bar{v}^2} \cdot \vec{x}$, the integral can be extended over the coordinate to infinity. Note that, alternatively, one can consider oscillations as a function of time and integrate over space, which leads to a similar expression for the decoherence term \cite{Akhmedov:2017mcc}.}
\begin{equation}\label{eq:rhot}
\rho_{jk}  (\vec{x}) \equiv \int  \mathrm dt \rho_{jk} (t,\vec{x}),
\end{equation}
the Gaussian integration gives the averaged density matrix, as a product of three factors 
\begin{equation}\label{eq:rhofsp}
\rho_{jk} (\vec{x})= A_{jk}^{\alpha} ~ \rho^{osc}_{jk}(\vec{x}) ~ \rho_{jk}^{damp}(\vec{x}).
\end{equation}
The first exponential term is
\begin{equation}\label{eq:Aconst}
A_{jk}^{\alpha} = U^*_{\alpha j} U_{\alpha k} \frac{\sqrt{2}}{ 2 \pi \sigma_x^2 \bar{v}} \exp{\left[ - \frac{(E_{jk} \sigma_x)^2}{\bar{v}^2} \right]},
\end{equation}
with $\bar{v} = \sqrt{v_j^2+v_k^2}$, has no influence on oscillations. 
The second one reads
\begin{equation}\label{eq:osc}
\rho^{osc}_{jk}(\vec{x}) = \exp{\left[ i\Big( \vec{p}_{jk} - \frac{2 E_{jk} \vec{v}_g}{\bar{v}^2}\Big) \vec{x} \right]}, 
\end{equation}
\noindent
which is the oscillation term, with the additional factor  $2 E_{jk} \vec{v}_g \bar{v}^{-2}$, $\vec{v}_g \equiv \frac{1}{2} (\vec{v}_j+\vec{v}_k)$ being the average group velocity, arising from the WP description. 

The last factor in Eq.(\ref{eq:rhofsp}) is the damping term 
\begin{equation}\label{eq:rhodamp}
\rho_{jk}^{damp}(\vec{x}) = \exp{\left[ - \frac{(\vec{v}_j-\vec{v}_k)^2 x^2}{4\sigma_x^2\bar{v}^2} \right]},
\end{equation}
that is responsible for decoherence\footnote{Note that both the averaged and the unaveraged density matrices have the same damping factor as a function of time (when integration over distance is performed instead of the one over time) \cite{Akhmedov:2017mcc}. }. 
From this expression, the coherence length $L_{coh}$ between the Gaussian $jk$ mass eigenstate WPs is 
\begin{equation}\label{eq:lcohf}
L_{coh} = \frac{2 \sigma_x \bar{v}}{\left|\vec{v}_j- \vec{v}_k \right|} \simeq \frac{4 \sqrt{2} E^2}{\left| \Delta m^2_{jk} \right|} \sigma_x.
\end{equation}
One can see that Eq.(\ref{eq:lcohf}) agrees, up to a factor, with the coherence length Eq.(\ref{e:Lcoh})
from the heuristic argument. Numerically, the coherence length is rather short, as it ranges between $11$ km and $83$ km, with a width $\sigma_x $ between $4 \times 10^{-12}$ cm and $10^{-11}$ cm and an energy $E$ between $11$ MeV and $20$ MeV. If the coherence length remains of
the same order of magnitude in the presence of matter and self-interactions, this could influence neutrino flavor conversion mechanisms. 

\section{Neutrino WP decoherence in curved spacetime}
\label{curved spacetime}
\noindent
In curved spacetime, proper times are measureable quantities. Therefore, a coherence proper time 
 appears more suitable than a coherence length to quantify neutrino WP decoherence, in presence of strong gravitational fields. We first present some kinematical arguments and then the derivation of the coherence proper time in the density matrix approach.

A neutrino flavor state, produced at the spacetime point $P$ $( t_P, \vec{x}_P ) $, is described by
\begin{equation}
 | \nu_{\alpha}  ( P ) \rangle =   U^*_{\alpha j} |  \nu_j  ( P )\rangle .
\end{equation}
The $j$th-mass eigenstate evolves from the production point $P$ to a "detection" point $D_j$ $( t_{D_j}, \vec{x}_{D_j} ) $ according to
\begin{equation}
  | \nu_j  (P,D_j) \rangle = e^{-i\phi_j ( P, D_j )}  |  \nu_j (P) \rangle,
\end{equation}
where the covariant form of the quantum mechanical phase is given by \cite{Stodolsky:1986dx}
\begin{equation}
\phi_j ( P, D_j ) = \int_P^{D_j}  p_\mu^{( j )} \mathrm d x^\mu.
\label{eq:curvedphase}
\end{equation}
The quadrivector $p_\mu^{( j )}$ is the canonical conjugate momentum to the coordinate $x^\mu$ 
\begin{equation}
p_\mu^{( j )} = m_j g_{\mu \nu} \frac{\mathrm dx^\nu}{\mathrm ds},
\end{equation}
with $g_{\mu\nu}$ being the metric tensor and $\mathrm ds$ the line element along the trajectory of the $j$th neutrino mass eigenstate.

\begin{figure}
\begin{center}
\includegraphics[width=.3\textwidth]{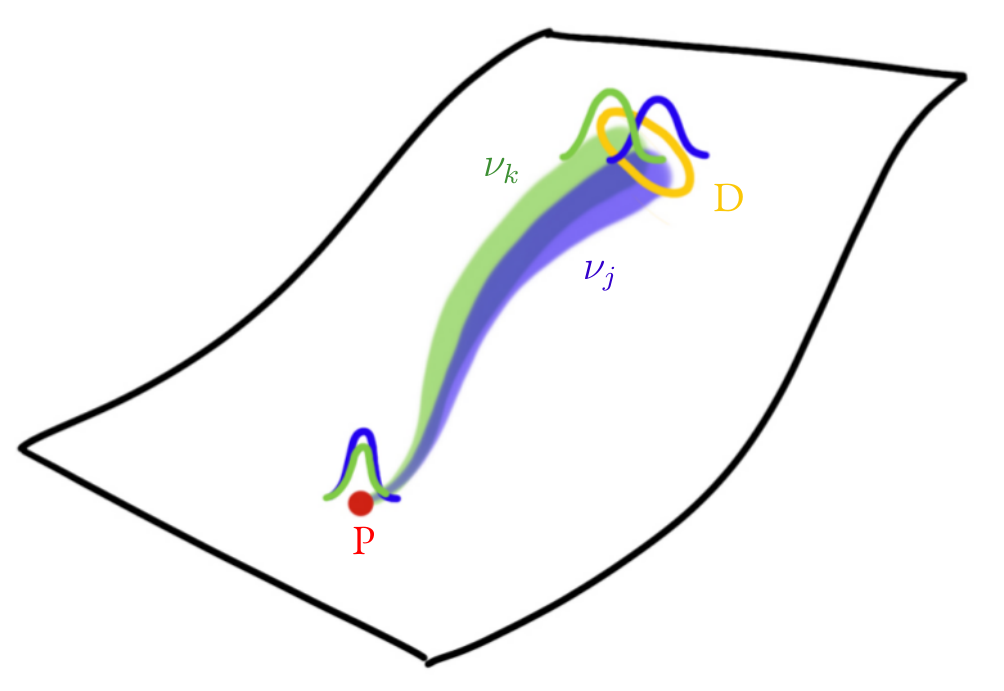}
\caption{Artistic drawing of a neutrino propagating from a production point P to "detection" point D
where the WPs can still interfere. Each mass eigenstate WP follows a trajectory close to null-geodesics. The coloured widths picture the distribution of trajectories due to the WP finite extension. }
\label{fig1}
\end{center}
\end{figure}

In presence of strong gravitational fields, the phase differences 
are usually calculated along null-geodesics (see e.g. \cite{Fornengo:1996ef,Cardall:1996cd}). In order to evaluate the decoherence of the neutrino WPs in curved spacetime 
we assume that the ensemble of trajectories for each mass eigenstate
is close to null-geodesics. To determine the coherence proper time,  
a spacetime point $D$ $( t_D, \vec{x}_D ) $ is considered, at which the WPs can still interefere (Figure \ref{fig1}).

\subsection{Neutrino trajectories in the Schwarzschild metric}
The Schwarzschild metric for a static gravitational field with spherical symmetry is
\begin{equation}
\mathrm ds^2 = - B(r) \mathrm dt^2 + \frac{1}{B(r)} \mathrm dr^2+ r^2 \mathrm d\theta^2 + r^2 \sin^2{\theta} d\varphi^2,
\label{eq:dsschwarz}
\end{equation}
where $\left(t, r, \theta, \varphi \right)$ are time, radial distance and angular coordinates and 
\begin{equation}
B(r) = 1-\frac{r_s}{r}, ~~~~~~ r_s = 2M,
\end{equation}
with $r_s$ the Schwarzschild radius and $M$ the mass of the central object. 
Since the gravitational field is spherically symmetric, the neutrino trajectories are confined to a plane. We choose to work in the plane $\theta = \frac{\pi}{2}$. 
The relevant components of $p_\mu^{(j)}$ are 
\begin{eqnarray}
p_t^{(j)} & = & -m_j B(\vec{r}) \frac{\mathrm dt}{\mathrm ds}, \label{eq:ptschwarz} \\
p_r^{(j)} & = & \frac{m_j}{B(\vec{r})} \frac{\mathrm dr}{\mathrm ds}, \label{eq:prschwarz}\\
p_\varphi^{(j)} & = & m_j r^2 \frac{\mathrm d\varphi}{\mathrm ds} \label{eq:pfschwarz}.
\end{eqnarray}
They are related by the mass on-shell relation 
\begin{equation}\label{eq:onshell}
p_\mu^{(j)} p^{(j)\mu} = -m_j^2 .
\end{equation} 
Since the metric tensor $g_{\mu \nu}$ does not depend on $t$ and $\varphi$, the canonical momentum components 
\begin{equation}
E_j ( \vec{p}) \equiv -p_t^{(j)}, ~~~~  J_j( \vec{p})= p_\varphi^{(j)}, 
\end{equation}
are constants of motion. 
They correspond to the energy and the angular momentum of the $j$th mass eigenstate seen by an observer at $r= +\infty$ and therefore
differ from those measured by an observer at $D$, or at the production point $P$. 
Obviously the local energy, measured by an observer at rest at a given spacetime point, can be related to $E_j$ through a transformation between the two frames. 

The phase argument in \eqref{eq:curvedphase} can be developed as
\begin{equation}
p_\mu^{(j)} \mathrm d x^\mu = - E_j (\vec{p}) \mathrm dt + \frac{m_j}{B(r)} \Big( \frac{\mathrm dr }{\mathrm ds} \Big) \mathrm dr + J_j (\vec{p}) \mathrm d\varphi.
\label{eq:curvedpmudxmu}
\end{equation}
We consider the case of radial propagation\footnote{Note that e.g. in Ref.\cite{Fornengo:1996ef}, the cases of radial and of non-radial propagation are considered.}, i.e. $\mathrm d\varphi=0$, for which the mass on-shell relation Eq.(\ref{eq:onshell}) becomes
\begin{equation}\label{eq:1}
- B(r) \left( \frac{\mathrm dt}{\mathrm ds} \right)^2 + \frac{1}{B(r)} \left( \frac{\mathrm dr}{\mathrm ds} \right)^2 = -1.
\end{equation}
By using \eqref{eq:ptschwarz} along with $p_t^{( j )} = -E_j \left( \vec{p} \right)$, Eq.\eqref{eq:1} reads
\begin{equation}
\frac{1}{B(r)} \left( \frac{\mathrm dr}{\mathrm ds} \right)^2 = -1 + \frac{E_j^2( \vec{p} )}{m_j^2} \frac{1}{B(r)},
\end{equation}
which gives
\begin{equation}
\frac{\mathrm dr}{\mathrm ds} = \sqrt{\frac{E_j^2 ( \vec{p} )}{m_j^2}-B(r) },
\label{eq:radialprop}
\end{equation}
assuming that neutrinos are propagating outwards. 
We now introduce general kinematical arguments that will be used to estimate the coherence proper time $\tau_{coh}$.

\subsection{A kinematical argument}
\noindent
A clock at the "detection" point D measures the time delay between the arrival of the WPs of the $j$th and $k$th mass eigenstates propagating along radial geodesics, from P to  D. By combining Eqs.\eqref{eq:ptschwarz} and \eqref{eq:radialprop} one gets
\begin{equation}
\frac{\mathrm dr}{\mathrm dt} = B(r)\sqrt{1 - \frac{m_{j}^2 B(r)}{E^2_{j}(\vec{p})} }.
\label{eq:ddrdt}
\end{equation}
By inverting this relation, we get that the $j$th mass eigenstate WP reaches D at 
the coordinate time 
\begin{equation}
t^{j}_{PD} = \int_{r_P}^{r_D} \frac{dr}{B(r)}{\left[1 - \frac{m_{j}^2 B(r)}{E^2_{j}(\vec{p})}\right]}^{-1/2}.
\label{eq:tpd}
\end{equation}
In the limit $m_{j}^2 B(r)/E^2_{j}(\vec{p}) \ll 1$, one finds at first order
\begin{equation}
t^{j}_{PD} =  \frac{m_{j}^2}{2E^2_{j}(\vec{p})}r_{PD} + b_{PD},
\label{eq:tj}
\end{equation}
where the second term is 
\begin{equation}
b_{PD} = \int_{r_P}^{r_D} \frac{dr}{B(r)} = r_{PD} + r_s \ln(\frac{r_D-r_s}{r_P - r_s}),
\label{eq:bpd}
\end{equation}
with $r_{PD} = r_D - r_P$.
Therefore, from Eqs.(\ref{eq:tj}-\ref{eq:bpd}), one gets for the coordinate time delay between the two WPs
at D 
\begin{equation}
t_{PD}^{jk} = \left[\frac{m_{j}^2}{2E^2_{j}(\vec{p})} - \frac{m_{k}^2}{2E^2_{k}(\vec{q})}\right]r_{PD}.
\label{eq:tjk}
\end{equation}
Now, an observer in D will measure a proper time
\begin{equation}
\tau_D = t_D \sqrt{B(r_D)},
\label{eq:tau}
\end{equation}
where we have introduced the WP dispertion in the coordinate time  $\sigma_t$.
Combining \eqref{eq:tjk} and \eqref{eq:tau} gives the difference between the $j$th and 
$k$th WPs proper times in D
\begin{equation}
\tau_{D}^{jk} = \sqrt{B(r_D)} \left[\frac{m_{j}^2}{2E^2_{j}(\vec{p})} - \frac{m_{k}^2}{2E^2_{k}(\vec{q})}\right]   r_{PD}.
\label{eq:dtau}
\end{equation}

In analogy with the coherence length \eqref{e:gwp} in flat spacetime, one can define a coherence proper time $\tau_{coh}$ at which the difference in the proper times at D satisfies the following relation
\begin{equation}
 \tau = {\tau}_{coh} ~~~~\mathrm{if}~~~~\tau^{jk}_{D} = \sigma_t \sqrt{B(r_D)}.
\label{eq:cohpropt}
\end{equation}
Therefore, from \eqref{eq:dtau}-\eqref{eq:cohpropt}, one gets for the corresponding coordinate distance
\begin{equation}
r^{coh}_{PD} = {2 E^2 \over{\Delta m^2_{jk}}} \sigma_x,  
\label{eq:discoh}
\end{equation}
with the assumptions that $\sigma_x \simeq \sigma_t$ and $E_{j}(\vec{p}) \simeq E_k (\vec{q}) \simeq E$.
This relation is general\footnote{Note however that $\sigma_x \simeq \sigma_t$.} and will be used as a comparison for the results based on the density matric approach. 

\subsection{The density matrix approach}
\noindent
Let us now consider the covariant phase Eq.\eqref{eq:curvedphase}. From Eqs. \eqref{eq:curvedpmudxmu} and \eqref{eq:radialprop} one gets for the integral argument, in the case of radial propagation  
\begin{equation}
p_\mu^{( j )} \mathrm d x^\mu = - E_j( \vec{p} ) \mathrm dt + \frac{1}{B(r)} \sqrt{E_j ( \vec{p} )^2-B(r)  m_j^2} \mathrm dr.
\label{eq:pmudxmuj}
\end{equation}
Neutrinos are assumed to be relativistic at infinity, i.e. $m_j(E_j)^{-1} \ll 1$, which ensures that the conditions is satisfied everywhere on their trajectory\footnote{Note that this is not necessarily the case if neutrinos are considered to be relativistic at the source \cite{Fornengo:1996ef}.}. Equation \eqref{eq:pmudxmuj} becomes
\begin{equation}\label{eq:pdx}
p_\mu^{( j )} \mathrm d x^\mu = - E_j ( \vec{p} )\mathrm dt + \frac{1}{B(r)} \left[  E_j ( \vec{p} )- \frac{m_j^2}{2E_j ( \vec{p} )} B(r) \right] \mathrm dr.
\end{equation}
From Eqs.\eqref{eq:curvedphase}, \eqref{eq:bpd} and \eqref{eq:pdx}, the covariant phase reads
\begin{multline}
\phi_j ( P, D ; \vec{p} ) =  - E_j ( \vec{p} )( t_{PD} 
-  b_{PD}) - 
\frac{m_j^2}{2 E_j ( \vec{p})} r_{PD},
\end{multline}
where $t_{PD} = t_D - t_P$. Consequently, the phase difference $\phi_{kj} = \phi_k-\phi_j$ is
\begin{multline}
\phi_{kj} ( P, D ; \vec{p}, \vec{q} )   =  ( E_j ( \vec{p} ) - E_k ( \vec{q} ) )  ( t_{PD} - b_{PD})\\ 
 +  \left[  \frac{m_j^2}{2 E_j ( \vec{p})} - \frac{m_k^2}{2 E_k ( \vec{q})} \right]r_{PD},
\end{multline}
and can be written as
\begin{multline}\label{eq:phijkdev}
\phi_{kj} ( P, D ; \vec{p}, \vec{q} ) = E_{jk}  ( t_{PD} - b_{PD}) 
 + \left(  \frac{m_j^2}{2 E_j } - \frac{m_k^2}{2 E_k} \right)r_{PD} \\ 
+   \vec{v}_{j}  ( \vec{p} - \vec{p}_{j} ) \left[t_{PD} - \lambda_{j} r_{PD}  \right]
-   \vec{v}_{k}  ( \vec{q} - \vec{p}_{k} ) \left[t_{PD} - \lambda_{k} r_{PD} \right],
\end{multline}
by using the first-order expansion \eqref{eq:expansionE}, with the notation\footnote{Similarly for $\lambda_k$. The explicit dependence of $\lambda$ on PD is not shown to simplify notations.} 
\begin{equation}
\lambda_j  = \frac{m_j^2}{2 E_j^2} r_{PD} + b_{PD}.
\end{equation}

Let us now introduce one-body density matrices \eqref{eq:exprhojk} describing the neutrino mass eigenstates as (non-covariant) Gaussian WPs of width $\sigma_p$ 
\begin{multline}
\rho_{jk} ( P, D ) = \int_{\vec{p}}\int_{\vec{q}} \exp{\left[- i\phi_{kj} ( P,D ; \vec{p}, \vec{q} )  \right]} \\
 \exp{\left[- \frac{( \vec{p}-\vec{p}_j )^2}{4\sigma_p^2} - \frac{( \vec{q}-\vec{p}_k )^2}{4\sigma_p^2} \right]}. 
\label{eq:exprhojkxcurved}
\end{multline}
By using \eqref{eq:phijkdev} the Gaussian integrals can be performed giving the following expression for the elements of the one-body density matrix 
\begin{multline}
\rho_{jk} ( P, D ) = 
 \exp{ \{  - \sigma_p^2 \left[ v_k^2 ( t_{PD} - \lambda_k )^2   + v_j^2 ( t_{PD} - \lambda_j )^2 \right]  \}} \\
\times N^{\alpha}_{jk}\exp{ \left[ - i  E_{jk}  ( t_{PD}  - b_{PD} ) + i \left( \frac{m_j^2}{2 E_j } - \frac{m_k^2}{2 E_k} \right) r_{PD}  \right]},
\label{eq:exprhojktxcurved}
\end{multline}
with \eqref{eq:factor} for the normalisation factor $N^{\alpha}_{jk}$.

We introduce the density matrix integrated over coordinate time
\begin{equation}
\rho_{jk} ( r_P, r_D ) = \int \mathrm dt \rho_{jk} ( P, D ), 
\end{equation}
and compute the Gaussian integral, which gives
\begin{equation}
\rho_{jk} ( r_P, r_D )= A_{jk}^{\alpha} ~  \rho^{osc}_{jk} ( r_P, r_D )  ~ \rho^{damp}_{jk}( r_P, r_D ),
\end{equation}
to be compared with the flat spacetime expression \eqref{eq:rhofsp}. 
The first exponential term does not depend on $r_{PD}$ and is the same as in flat spacetime \eqref{eq:Aconst}. It has no influence on neutrino propagation.
The second exponential term, that generates neutrino oscillations, is
\begin{multline}
 \rho^{osc}_{jk} (r_P, r_D) =
\exp{\left[i\left(  \frac{m_j^2}{2 E_j } - \frac{m_k^2}{2 E_k} \right) r_{PD} \right] } \\
\times \exp{ \left[- i  \frac{E_{jk}}{\bar{v}^2} \left( v_k^2 \frac{m_k^2}{2E_k^2} +   v_j^2 \frac{m_j^2}{2E_j^2 } \right)r_{PD} \right] },
\end{multline}
where, in the Schwarzschild metric, $r_{PD}$  does not represent a physical distance.
The third and last factor is the damping term responsible for decoherence
\begin{equation}
\rho^{damp}_{jk}( r_P, r_D ) = \exp{\left[-\frac{({v}_j {v}_k r_{PD})^2 }{4 \sigma_x^2 \bar{v}^2}  
\left( \frac{m_k^2}{2E_k^2}- \frac{m_j^2}{2E_j^2}\right)^2 \right]},
\label{eq:exprhojkxcurvedint}
\end{equation}
which becomes at first order in $m_j/E$ 
\begin{equation}
\rho^{damp}_{jk}( r_P, r_D ) \simeq \exp{\left[-\frac{ \Delta m_{jk}^4 r_{PD}^2}{32 \sigma_x^2 E^4} \right]}.
\label{eq:dampingterm}
\end{equation}
In the flat spacetime limit, $r_{PD}$ becomes the physical distance travelled by neutrinos and this term reduces to the damping term \eqref{eq:rhodamp} with $r_{PD}$ the coherence length \eqref{eq:lcohf}. However, if $r_s$ is non-null, $r_{PD}$ does not represent a physical distance, while $E$ is not the local energy of the neutrinos but rather the energy at infinity. 

In analogy with the flat spacetime case, from Eq.\eqref{eq:dampingterm} one can define a coherence coordinate distance $r^{coh}_{PD}$ at which the density matrix gets suppressed by $e^{-1}$, namely
\begin{equation}
r^{coh}_{PD} = \frac{4 \sqrt{2} \sigma_x E^2}{\Delta m_{jk}^2}.
\label{eq:rcoh}
\end{equation}
Note that formally, the expression for $r^{coh}_{PD}$ is the same as the coherence length in flat spacetime \eqref{eq:lcohf}.
The comparison with \eqref{eq:discoh} shows the coherent coordinate distance $r^{coh}_{PD}$ agrees
with the expression \eqref{eq:discoh} from kinematical arguments up to $2 \sqrt{2}$\footnote{Note that the heuristic coherence length \eqref{e:Lcoh} also differs from \eqref{eq:lcohf}, derived from the density approach, by the same factor, which comes from the shape chosen for the WPs.}.

\subsection{The coherence proper time}
\noindent
We now use kinematical arguments to relate the derived coherence coordinate distance to a coherence proper time. We start by noticing that the null-geodesics can be used to express the travelled distance as a function of the coordinate time to go from P to D. By using Eqs.\eqref{eq:1} and \eqref{eq:tj}, for null-geodesics, one has 
\begin{equation}
t_{coh}^{travel} = b_{PD},
\label{eq:tcoh}
\end{equation}
and from \eqref{eq:tau}, the coherence proper time is 
\begin{equation}\label{eq:taucoh}
\tau_{coh} = \sqrt{B(r^{coh}_D)}\left[r^{coh}_{PD} + r_s \ln \left(1 + \frac{r^{coh}_{PD}}{r_P - r_s}\right)\right],
\end{equation}
where $r^{coh}_D = r^{coh}_{PD} + r_P$.

\begin{figure}
\begin{center}
\includegraphics[width=.3\textwidth]{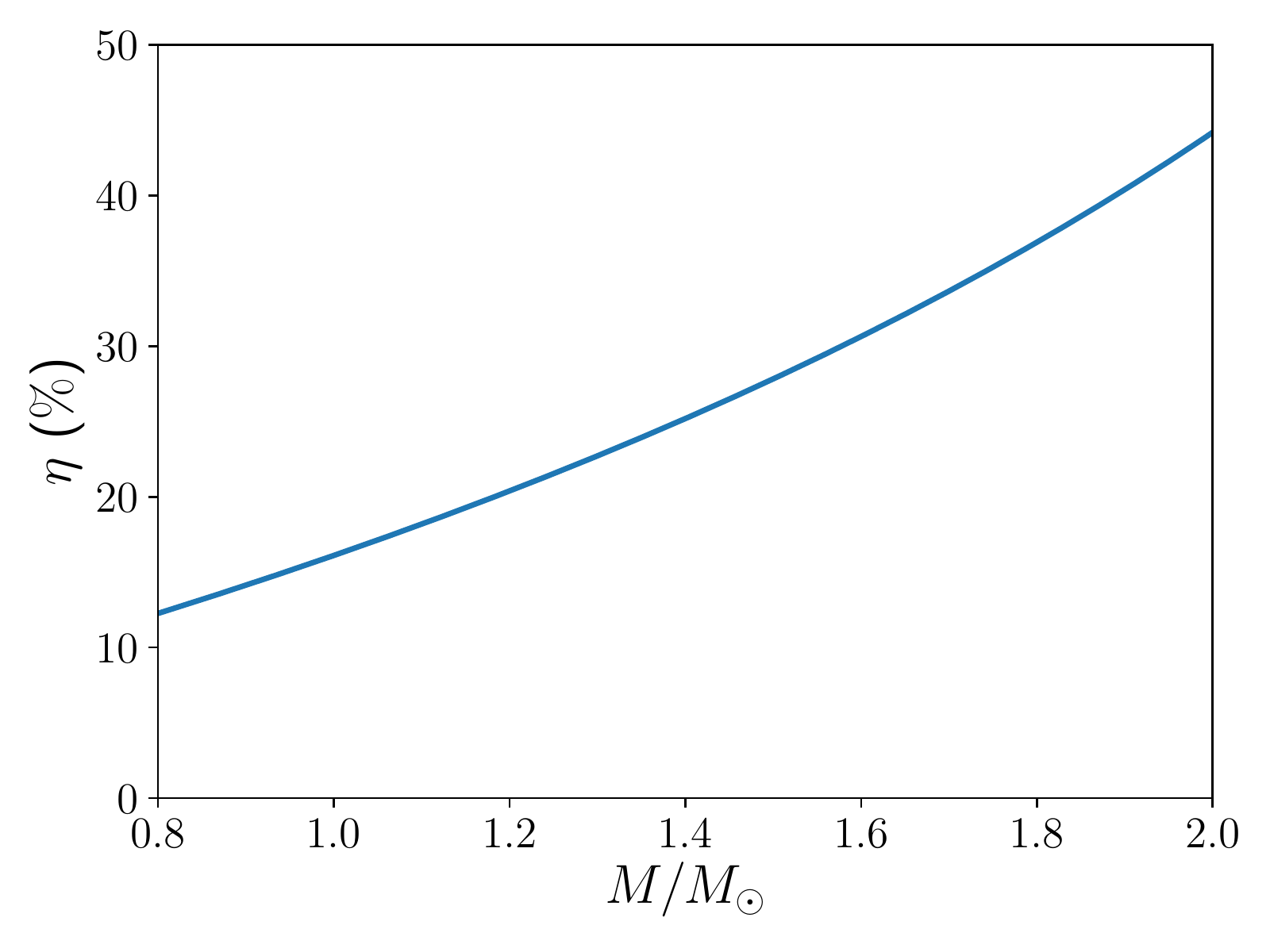}
\caption{Numerical estimates of the relative difference between the coherence proper time in curved spacetime and the coherence length in flat spacetime, as a function of the Schwarzschild mass $M$.}
\label{fig:numestlcoh}
\end{center}
\end{figure}

In order to show the impact of strong gravitational fields
on the coherence proper time, we present an estimate considering 
the case of a newly formed neutron star from a core-collapse supernova. Figure \ref{fig:numestlcoh} shows 
the relative difference between the coherence proper time \eqref{eq:taucoh} and the flat spacetime case \eqref{eq:lcohf}
\begin{equation}\label{eq:eta}
\eta = \frac{\tau_{coh} - L_{coh}}{L_{coh}}(\%),
\end{equation}
as a function of the Schwarzschild mass $M \in [0.8,2]~M_\odot$. 
Neutrinos are emitted at a neutrinosphere of radius $r_P = R_\nu = 10$ km and a typical energy of $E=11$ MeV with $r^{coh}_{PD}$ given by Eq.\eqref{eq:rcoh}. For the WP width we take $\sigma_x \approx 4\times 10^{-12}$ cm \cite{Kersten:2015kio,Akhmedov:2017mcc}. From Figure \ref{fig:numestlcoh}, one can see that the influence of the gravitational field is significant, being of several tens of percent, about $25 \%$ ($45 \%$) for $1.4 M_{\odot}$ ($2 M_{\odot}$). 

\section{Conclusions}
\label{num}
\noindent
In the present manuscript we have explored neutrino decoherence from WP separation in curved spacetime. To this aim we have extended the WP density matrix formalism used in the context of flat spacetime. We have performed our calculations in the static and spherically symmetric Schwarzschild metric, considering the WPs travel along radial geodesics. We have derived the coherence radial coordinate at a distance from the production point where the WPs still interfere and shown that it is consistent with the one obtained from kinematical arguments. We have then related it to the coherence proper time and provided a numerical estimate showing that the impact of strong gravitation fields on the coherence proper time can be sizable.

This is a first step in the investigation of decoherence effects in presence of strong gravitational fields. Future studies should address the role on the coherence proper time of neutrino interactions with matter and neutrino self-interactions, outside the compact central object. 
In particular, for adiabatic evolution, a similar procedure could be used based on the matter eigenstate basis. These investigations are necessary to assess if WP decoherence suppresses flavor evolution and
its potential impact on the supernova dynamics, $r$-process nucleosynthesis as well as future supernova neutrino observations.  

\vspace{.3cm}
The authors are grateful to Nathalie Deruelle for pointing out the kinematical arguments
they have used to define a coherent proper time.
They would like to thank Gaetano Lambiase and Carlo Giunti for their comments and acknowledge support from ”Gravitation et physique fondamentale” (GPHYS) and "Physique Fondamentale et Ondes 
Gravitationnelles" (PhyFOG) of the Observatoire de Paris.


\vspace{.5cm}
\begin{thebibliography}{00}

\bibitem{Fukuda:1998mi} 
  Y.~Fukuda {\it et al.} [Super-Kamiokande Collaboration],
  Phys.\ Rev.\ Lett.\  {\bf 81}, 1562 (1998)
  [hep-ex/9807003].

\bibitem{Ahmad:2002jz} 
  Q.~R.~Ahmad {\it et al.} [SNO Collaboration],
  Phys.\ Rev.\ Lett.\  {\bf 89}, 011301 (2002)
  [nucl-ex/0204008].


\bibitem{Capozzi:2018ubv} 
  F.~Capozzi, E.~Lisi, A.~Marrone and A.~Palazzo,
  Prog.\ Part.\ Nucl.\ Phys.\  {\bf 102}, 48 (2018)
  [arXiv:1804.09678 [hep-ph]].

\bibitem{Nussinov:1976uw} 
  S.~Nussinov,
  Phys.\ Lett.\  {\bf 63B}, 201 (1976).


\bibitem{Giunti:2003ax} 
  C.~Giunti,
  Found.\ Phys.\ Lett.\  {\bf 17}, 103 (2004)
  [hep-ph/0302026].



\bibitem{Giunti:2007ry} 
  C.~Giunti and C.~W.~Kim,
  Oxford, UK: Univ. Pr. (2007) 710 p


\bibitem{An:2016pvi} 
  F.~P.~An {\it et al.} [Daya Bay Collaboration],
  Eur.\ Phys.\ J.\ C {\bf 77}, no. 9, 606 (2017)
  [arXiv:1608.01661].


  
\bibitem{Barenboim:2006xt} 
  G.~Barenboim, N.~E.~Mavromatos, S.~Sarkar and A.~Waldron-Lauda,
  Nucl.\ Phys.\ B {\bf 758}, 90 (2006)
  [hep-ph/0603028].


\bibitem{Fogli:2007tx} 
  G.~L.~Fogli, E.~Lisi, A.~Marrone, D.~Montanino and A.~Palazzo,
  Phys.\ Rev.\ D {\bf 76}, 033006 (2007)
  [arXiv:0704.2568].



\bibitem{Kersten:2015kio} 
  J.~Kersten and A.~Y.~Smirnov,
  Eur.\ Phys.\ J.\ C {\bf 76}, no. 6, 339 (2016)
  [arXiv:1512.09068].
 
\bibitem{Akhmedov:2017mcc} 
  E.~Akhmedov, J.~Kopp and M.~Lindner,
  JCAP {\bf 1709}, no. 09, 017 (2017)
  [arXiv:1702.08338].

\bibitem{Caballero:2011dw} 
  O.~L.~Caballero, G.~C.~McLaughlin and R.~Surman,
  Astrophys.\ J.\  {\bf 745}, 170 (2012)
  [arXiv:1105.6371].

\bibitem{Deaton:2018ser} 
  M.~B.~Deaton {\it et al.},
  Phys.\ Rev.\ D {\bf 98}, no. 10, 103014 (2018)
  [arXiv:1806.10255].

\bibitem{Fornengo:1996ef} 
  N.~Fornengo, C.~Giunti, C.~W.~Kim and J.~Song,
  Phys.\ Rev.\ D {\bf 56}, 1895 (1997)
  [hep-ph/9611231].
  
\bibitem{MosqueraCuesta:2017iln} 
  H.~J.~Mosquera Cuesta, G.~Lambiase and J.~P.~Pereira,
  Phys.\ Rev.\ D {\bf 95}, no. 2, 025011 (2017)
  [arXiv:1701.00431].

\bibitem{Cardall:1996cd} 
  C.~Y.~Cardall and G.~M.~Fuller,
  Phys.\ Rev.\ D {\bf 55}, 7960 (1997)
  [hep-ph/9610494].
  
  
\bibitem{Yang:2017asl} 
  Y.~Yang and J.~P.~Kneller,
  Phys.\ Rev.\ D {\bf 96}, no. 2, 023009 (2017)
  [arXiv:1705.09723].
  
\bibitem{Duan:2010bg} 
  H.~Duan, G.~M.~Fuller and Y.~Z.~Qian,
  Ann.\ Rev.\ Nucl.\ Part.\ Sci.\  {\bf 60}, 569 (2010)
  doi:10.1146/annurev.nucl.012809.104524
  [arXiv:1001.2799 [hep-ph]].
  

\bibitem{Tanabashi:2018oca} 
  M.~Tanabashi {\it et al.} [Particle Data Group],
  Phys.\ Rev.\ D {\bf 98}, no. 3, 030001 (2018).
  doi:10.1103/PhysRevD.98.030001
  
\bibitem{Volpe:2013jgr} 
  C.~Volpe, D.~V\"a\"an\"anen and C.~Espinoza,
  Phys.\ Rev.\ D {\bf 87}, no. 11, 113010 (2013)
  [arXiv:1302.2374].
  
\bibitem{Naumov:2013uia} 
  D.~V.~Naumov,
  Phys.\ Part.\ Nucl.\ Lett.\  {\bf 10}, 642 (2013)
  [arXiv:1309.1717].


\bibitem{Stodolsky:1986dx} 
  L.~Stodolsky,
  Phys.\ Rev.\ D {\bf 36}, 2273 (1987).
  



\end{thebibliography}
\end{document}